\documentclass[12pt]{article}
\usepackage{amsfonts}
\usepackage{amsmath,mathrsfs,bm,amssymb,color}%,bbm}
\usepackage[width=15cm,height=8.5in]{geometry}
\usepackage[dvipdfmx]{hyperref}

\newtheorem{theorem}{Theorem}[section]
\newtheorem{definition}[theorem]{Definition}
\newtheorem{lemma}[theorem]{Lemma}
\newtheorem{proposition}[theorem]{Proposition}
\newtheorem{corollary}[theorem]{Corollary}
\newtheorem{remark}[theorem]{Remark}

\providecommand{\bc}[1]{\begin{corollary}\label{#1}}
\providecommand{\ec}{\end{corollary}}
\providecommand{\bp}[1]{\begin{proposition}\label{#1}}
\providecommand{\ep}{\end{proposition}}

\newenvironment{proof}[1][Proof]{{\sc #1.} }{\hfill $\Box$}

\newcommand{\ren}{\E}

\newcommand{\bt}[1]{\begin{theorem}\label{#1}}
\newcommand{\et}{\end{theorem}}
\newcommand{\bl}[1]{\begin{lemma}\label{#1}}
\newcommand{\el}{\end{lemma}}

\newcommand{\br}[1]{\begin{remark}\label{#1}}
\newcommand{\er}{\end{remark}}

\makeatletter
\@addtoreset{equation}{section}
\makeatother

\def\bbbone{{\mathchoice {\rm 1\mskip-4mu l} {\rm 1\mskip-4mu l}
{\rm 1\mskip-4.5mu l} {\rm 1\mskip-5mu l}}}
\def\one{\bbbone}

\newcommand{\akl}{A_{\eps}}

\renewcommand{\k}{\kappa}

\newcommand{\vz}{{\hat {\varrho}_0}}
\newcommand{\liml}{\lim_{\eps\downarrow 0}}

\newcommand{\ai}{A_{\infty}} 
\newcommand{\cp}{C_{\rm p}} 
\newcommand{\am}{C_\ast}
\newcommand{\suk}{\sup_{|k|\geq \la}} 
 
\newcommand{\GGG}{(H_\La-\El+\omega(k))^{-\han}} 
\newcommand{\GGGG}{(H_\La-\El+\omega(k))\f}
\newcommand{\FF}{(H_0+ \one)^{-\han}} 

\newcommand{\grr}{\gr(\eps')}
\newcommand{\gz}{H_0}
\newcommand{\ckl}{c_{\la}} 
\newcommand{\vkl}{V_{\la}} 
\newcommand{\HL}{H_\eps}

\newcommand{\El}{E_{\eps}(g^2)} 
\newcommand{\E}{E_{\eps}^{\rm ren}}

\newcommand{\eel}{E_{\la}^\ast} 
\newcommand{\Ei}{E_{\infty}}
\newcommand{\at}{g^2}
\newcommand{\mm}{|g|} 

\providecommand{\FFF}{(H_0+\one)^{1/2}}
\newcommand{\m}{\left\|{\vpl}/{\sqrt\omega}\right\|}%\bk\|} 
\providecommand{\epp}{E_{\rm p}}
\newcommand{\pa}{H_{\rm p}}
\newcommand{\eht}{e^{-t\HL }}

\newcommand{\kp}{\one_{|k|\geq\la}}
\newcommand{\Ebb}{{\mathbb E}}

\providecommand{\eq}[1]{\begin{equation}\label{#1}}
\providecommand{\en}{\end{equation}}

\renewcommand{\d}{\displaystyle}

\providecommand{\qed}{\hfill {$\Box$}\par\medskip}
\providecommand{\BR}{{\mathbb{R}^3 }}

\providecommand{\bi}
{\begin{itemize}}
\providecommand{\ei}{\end{itemize} }
\providecommand{\CC}{{\mathbb{C}}}
\providecommand{\RR}{\mathbb{R}}

\providecommand{\kak}[1]{(\ref{#1})}
\providecommand{\LR}{{L^2(\BR)}}

\providecommand{\ZZZ}{}

\providecommand{\fff}{\mathscr{F}}
\providecommand{\is}{\inf \s}

\renewcommand{\sp}{\Sigma_{\rm p}}%{\rm Spec}}

\providecommand{\f}{^{-1}}
\providecommand{\lk}{\left(}
\providecommand{\rk}{\right)}
\providecommand{\lkk}{\left\{}
\providecommand{\rkk}{\right\}}

\providecommand{\add}{a^\ast}

\providecommand{\ov}[1]{\overline{#1}}
\providecommand{\hf}{H_{\rm f}}
\providecommand{\pf}{{P_{\rm f}}}

\providecommand{\gr}{\varphi_{\rm g}}
\providecommand{\grt}{\varphi_{\rm g}^T}

\providecommand{\half}{\frac{1}{2}}
\providecommand{\han}{{1/2}}

\providecommand{\he}{H_\eps}

\providecommand{\hi}{H_{\rm I}}
\providecommand{\hp}{H_{\rm p}}

\providecommand{\la}{{\lambda}}
\providecommand{\La}{{\eps}}

\providecommand{\s}{\sigma}
\providecommand{\vp}{{\hat  \varphi}}
\providecommand{\vpl}{{\hat  \varphi_\eps}}
\providecommand{\vpe}{{\hat  \varphi_\eps}}

\providecommand{\non}{\nonumber}
\providecommand{\vp}{\mathop{\mathrm{{\varphi}}}\nolimits}

\makeatletter
\@addtoreset{equation}{section}
\makeatother

\providecommand {\s}  {\ensuremath {\mathcal{S}}}

\providecommand{\eps}{\varepsilon}

\newcounter {constant}
\newenvironment{constant}{\refstepcounter{constant} }{}

\makeatletter
\@addtoreset{equation}{section}
\makeatother
\title
{Note on ultraviolet renormalization and ground state energy of the Nelson model}
\author{Fumio Hiroshima\\
Faculty of Mathematics, Kyushu University, \\ 
Fukuoka, 819-0395, Japan} 
\date{\today}
\begin{document}

\maketitle

%\tableofcontents
\begin{abstract}
Ultraviolet (UV) renormalization of the  Nelson model $H_\eps$ in quantum field theory is considered. 
E. Nelson proved that 
$\lim_{\eps\to 0} e^{-T(H_\eps-\E)}$ converges to $e^{-TH_{\rm ren}}$ in \cite{nel64}.  A relationship between a ultraviolet renormalization term $\E$ and the ground state energy $\El$ of the Hamiltonian with total momentum zero
$H_\eps(0)$ is studied by functional integrations.
Here $g$ denotes a coupling constant involved in $H_\eps(0)$. 
It can be derived from the formula 
$$\El=-\lim_{T\to\infty}\frac{1}{2T}\log(\one, e^{-2TH_\eps(0)}\one)$$
that 
$\E$ coincides with 
the coefficient of $g^2$ in the expansion of  $\El$ in $g^2$, i.e,
$\d \E=\lim_{g\to 0}\El/g^2$, and 
$\El-g^2\E$ converges as ultraviolet cutoff is removed. 
\end{abstract}

\section{The Nelson model}
In this paper we consider 
a relationship between 
a ultraviolet (UV) renormalization and the ground state energy of 
the Nelson model  in quantum field theory 
{\it by functional integrations}. 
The Nelson model describes an interaction system  
between a scalar bose field and particles governed by 
a Schr\"odinger operator with an external potential.  
%The Nelson Hamiltonian can be realized as a self-adjoint operator $H$ on a Hilbert space, on which UV cutoff is imposed to define $H$ as a self-adjoint operator. 
We prepare 
tools  used in this paper. 
The boson Fock space
$\fff$ over 
$\LR$ is defined by 
\eq{fff}
\fff=\bigoplus_{n=0}^\infty [\otimes_s^n \LR].
\en
Here 
$\otimes_s^n \LR$ describes $n$ fold symmetric tensor product of $\LR$ with $\otimes_s^0 \LR=\CC$. 
Let $\add (f)$ and $a(f)$, $f\in\LR$, be the creation operator and the annihilation operator, respectively, in $\fff$, which 
satisfy  $(\add(\bar f))^\ast=a(f)$ and 
canonical commutation relations:  
\eq{ccr}
[a(f),\add(g)]=(\bar f, g)\one,\quad
[a(f),a(g)]=0=[\add(f), \add(g)].
\en
Note that 
$(f,g)$ denotes the scalar product on $\LR$ and it is linear in $g$ and anti-linear in $f$. 
We also note that 
$f\mapsto \add(f)$ and $f\mapsto a(f)$ are  linear. 
Denote the dispersion relation by $\omega(k)=|k|$ . 
Then the  free field Hamiltonian 
$\hf$ of $\fff$ is then defined by the second quantization of 
$\omega$,
i.e., $ \hf =d\Gamma(\omega)=\int\omega(k) \add(k) a(k) dk$.
It satisfies that 
\eq{sasa}
e^{-it\hf} \add(f) e^{-it\hf}=\add(e^{-it\omega}f),\quad 
e^{-it\hf} a(f) e^{-it\hf}=a(e^{it\omega}f).
\en
Hence it follows that 
$$[\hf, a(f)]=-a(\omega f),\quad 
[\hf, \add(f)]=-\add(\omega f).
$$
Furthermore 
for the Fock vacuum $\one_\fff=1\oplus 0\oplus0\cdots\in\fff$, 
it follows that $\hf \one_\fff=0$.
\begin{definition}
The Nelson Hamiltonian $H$ is a self-adjoint operator acting in the Hilbert space 
$\LR\otimes\fff\cong L^2(\BR,\fff)$, 
which is given by 
\eq{nelson1}
H=(-\half \Delta+V)\otimes \one+\one\otimes\hf+g 
\phi,
\en
where $g\in\RR$ is a coupling constant,  
$V:\BR\to\RR$ an external potential,  the interaction is defined by $(\phi \Phi)(x)=\phi(x) \Phi(x)$ for a.e. $x\in\BR$ and 
the field operator $\phi(x)$ by 
\eq{int}
\phi(x)
  =  
\frac{1}{\sqrt 2}\lk
\add(\vp/\sqrt{\omega}e^{i(\cdot,x)})+
a(\widetilde{\vp}/\sqrt{\omega}e^{-i(\cdot,x)})
\rk
\en
\end{definition}
with $\widetilde{\vp}(k)={\vp}(-k)$. 
Let $H_0$ be the operator defined by $H$ with coupling constant $g$ replaced by $0$. 
We have to mention the self-adjointness of $H$. 
Suppose that 
\begin{align}
\label{sa}
\vp/\sqrt\omega,\vp/\omega\in\LR,\quad 
\vp(-k)=\ov{\vp(k)}.
\end{align}
Then the 
interaction 
$\hi$ is well defined, symmetric and  
infinitesimally 
$H_0$-bounded, i.e., for arbitrary $\eps>0$, 
there exists a $b_\eps>0$ such that 
$$\|\hi \Phi\|
\leq 
\eps  \|H_0\Phi\|+
b_\eps\|\Phi\|$$
for all $\Phi\in D(H_0)$. 
Thus
$H$ is self-adjoint on $D(H_0)$ by the Kato-Rellich theorem.
Throughout this paper we assume condition \kak{sa}.

\section{UV renormalization  and ground state energy}
A point charge limit of $H$,  
$\vp (k) \to \one $, 
is studied in \cite{nel64,nel64b} and recently 
in \cite{ghps12, ghl14,hir15}. 
Let $\la>0$ be a strictly positive infrared cutoff parameter 
and we fix it throughout. 
This assumption is used in the proof of Lemma \ref{main}.
Consider the cutoff function
\eq{sasa3}
\vpe(k)=e^{-\eps|k|^2/2}\kp,\quad \eps>0,
\en
and define the regularized Hamiltonian
by \begin{equation}
\he  = (-\half \Delta+V)\otimes \one  + \one\otimes \hf  + 
g \phi_\eps,\quad \eps>0,
\end{equation}
where 
$
\phi_\eps$ is 
defined 
by $\phi$ with $\vp$ replaced by 
$\vpe$.
Here $\eps>0$ is regarded as the UV cutoff parameter. 
Let 
\eq{sasa6}
\ren=-{g^2} \int_{|k|>\la}
\frac{e^{-\eps|k|^2}}{2\omega(k)}\beta(k) dk,
\en
where 
$\beta$  is given by 
\eq{beta}
\beta(k)=\frac{1}{\omega(k)+|k|^2/2}.
\en
Notice  that $\ren\to-\infty$ as $\eps\downarrow 0$.
E. Nelson proved the proposition below in \cite{nel64}.
\bp{nelson}
There exists a constant $C$ such that 
$H_{\rm ren}-\E>-C$ uniformly in $\eps$ 
and 
there exists a self-adjoint operator $H_{\rm ren}$ 
such that 
\eq{cp}
\d s-\lim_{\eps\downarrow 0} 
e^{-T(H_\eps-g^2 \ren )}= e^{-T H_{\rm ren}}.
\en
\ep
\proof Refer to see \cite{nel64}. 
\qed

Let $V=0$. Then $H_\eps$ is translation invariant, i.e., 
$$[H_\eps, P_{tot,\mu}]=0,\quad \mu=1,2,3,$$
where 
$P_{tot}$ is the total momentum defined by 
$P_{tot}=-i\nabla\otimes \one+\one\otimes \pf$. Here $\pf$ denotes the field momentum operator given by 
$\pf=d\Gamma(k)=\int k\add(k) a(k) dk $.
Thus $H_\eps$ can be decomposed as 
$H_\eps=\int_\BR^\oplus H_\eps(P) dP$, where 
\eq{p}
H_\eps(P)=\half (P-\pf)^2+\hf +g\phi_\eps(0)
\en
is a self-adjoint operator in $\fff$ for each $P\in\BR$. 
Let $\El=\is(H_\eps(0))$ be the bottom of the spectrum of 
the Nelson model with zero-total momentum, $P=0\in\BR$.
Suppose that formally $\El$ can be expanded in $g^2$ as 
$
\El=E_\eps(0)+a_2 g^2+a_4 g^4+\cdots,
$
and the ground state energy as
$
\gr=\one +g\phi_1+g^2\phi_2+\cdots.
$
Note that $E_\eps(0)=0$. 
Then from equation 
$
H_\eps(0) \gr=\El \gr$, 
we can derive the identity 
$\phi_1=-(\half \pf^2+\hf)^{-1}\phi_\eps(0)\one_\fff$ and 
$$a_2=- (\one_\fff, \phi_\eps(0)\phi_1)=-(\phi_\eps(0)\one_\fff, (\half \pf^2+\hf)^{-1}\phi_\eps(0)\one_\fff)=\E.$$
Hence 
$
a_2=\ren$ is derived. 
Furthermore 
we  expect 
 that $a_n$, $n\geq 4$, converges  as $\eps\downarrow 0$, 
 and hence 
\eq{hira}
\lim_{\eps\downarrow 0}|\El\ZZZ-g^2 \E|<\infty.
\en 
All the statements mentioned  above are however {\it informal}.  
In this paper we are concerned with these facts 
by functional integrations in non-pertubative way.  
We can  show 
\kak{hira} for arbitrary values of $g$, 
and 
$\d \lim_{g\to0}{\El\ZZZ}/{g^2}=
\E$
in Theorem \ref{g}. %main} and Corollary \ref{1}. 
\begin{remark}
{\rm In Proposition \ref{nelson},  
\kak{hira}
 is proven by an operator theory, 
we however prove this by applying functional integrations.
}\end{remark}

\section{Functional integrations}
Let $(B_t)_{t\in\RR}$ denote the $3$-dimensional Brownian motion 
on $C(\RR,\BR)$ with the Wiener measure $W$. 
$\Ebb [\cdots]$ denotes the expectation with respect to $W$ describing the Wiener measure starting 
from $0\in\BR$. 
\bl{fkf}
For $P\in\BR$, 
it 
follows that 
\eq{2}
( \one_\fff , e^{-2T\he(P)}  \one_\fff ) = 
 \Ebb 
\!\left[
e^{iP\cdot(B_{T}-B_{-T})}
e^{\frac{g^2}{2} S_\eps } \right],
\en
where
\eq{regint}
S_\eps =  \int _{-T}^T ds\int_{-T}^T dt
 W_\eps (B_t-B_s,t-s)
\en
and 
$W_\eps: \BR \times\RR \to \RR$ is given by 
\eq
{w}
W_\eps (x,t) =  
\int_{|k|\geq \la} \frac{e^{-\eps |k|^2} e^{-ik\cdot x} e^{-\omega(k)|t|}}{2\omega(k)}  dk.
\en
\el
\proof 
Refer to see \cite[Lemma 2.2]{hir15}.
\qed

Putting $P=0$ in Lemma \ref{fkf}, we have 
\eq{fkf1}
( \one_\fff , e^{-2T\he(0)}  \one_\fff ) = 
 \Ebb 
\!\left[e^{\frac{g^2}{2} S_\eps } \right].
\en
\bl{fkf2}
Let $\la>0$.
Then
\eq{fkf3}
\El=-\lim_{T\to\infty}\frac{1}{2T}\log(\one, e^{-2TH_\eps(0)} \one).
\en
In particular 
\eq{wa}
\El=-\lim_{T\to\infty}\frac{1}{2T}\log
 \Ebb 
\!\left[e^{\frac{g^2}{2} S_\eps } \right].
\en
\el
\proof
Since $\la>0$, it is shown that 
$H_\eps(0)$ has the unique ground state $\gr$ and it is strictly positive.  
See also Appendix. 
Hence $(\one, \gr)>0$. 
In particular $(\one, \gr)\not=0$. 
Thus \kak{fkf3}  follows.
\qed
%By Lemma \ref{fkf2}  we estimate $\d \frac{\El\ZZZ}{g^2}$. 
It can be seen that the pair potential  $W_\eps (B_t-B_s,t-s)$ 
is singular at the diagonal part $t=s$.
We shall remove the diagonal part by using the It\^o formula, which is done in \cite{ghl14}. 
We  introduce 
the function
\eq{rho}
\varrho_\eps(x,t) =  \int_{|k|\geq \la} \frac{e^{-\eps |k|^2} e^{-ik\cdot x-\omega(k)|t|}}{2 \omega(k)}
\beta(k) dk,\quad \eps\geq 0.
\en

\bl{wa2}
It follows that 
$$S_\eps=
S_\eps^{\rm ren}+4T\varrho_\eps(0,0), \quad \eps>0,$$
where 
\begin{align}
S_\eps^{\rm ren}
=
S_\eps^{OD}&+2 \int_{-T}^T \lk 
\int_{s}^{[s+\tau]}\nabla\varrho_\eps(B_t-B_s,t-s) \cdot dB_t \rk ds
  \non\\
&-2 \int_{-T}^T \varrho_\eps(B_{[s+\tau]}-B_s,[s+\tau]-s)ds.
\end{align}
Here $0<\tau<T$ is an arbitrary number, and $[t]=-T\vee t\wedge T$, 
$S_\eps^{OD}$ denotes the off-diagonal part 
given by 
$$S_\eps^{OD}=
2\int_{-T}^T ds\int_{[s+\tau]}^T W_\eps(B_t-B_s,t-s) dt$$
and 
the integrand is 
\begin{align*}
\nabla_\mu  \varrho_\eps(X,t)=\int_{|k|\geq \la}
\frac{-ik_\mu  e^{-ikX}e^{-|t|\omega(k)}e^{-\eps|k|^2}}{2\omega(k)}\beta(k)  dk.
\end{align*}
\el
\proof
It is shown 
by 
the It\^o formula  that
\eq{ito}
\int_{s}^{S} W_\eps (B_t-B_s,t-s) dt \\
= \varrho_\eps(0,0) -\varrho_\eps(B_{S}-B_s,S-s) +
\int_{s}^{S} \nabla\varrho_\eps(B_t-B_s,t-s)\cdot dB_t.
\en
Then the lemma follows directly. 
\qed

 $-\varrho_\eps(0,0)$ can be regarded as the diagonal part of $W_\eps$ and turns to be a renormalization term. 
I.e., we have the lemma below.
\bl{g}
Let $\eps>0$. 
Then 
$\ren =-\varrho_\eps(0,0)$.
\el

\bl{muri}
There exist constants $b>0$ and $c>0$  independent of $g$ 
such that 
for all $\eps>0 $,
\eq{mmm2}
\Ebb \left[
e^{\frac{g^2}{2}S_\eps^{\rm ren} }
\right]
 \leq e^{b(c+g^4T+g^2 \log T)+ c(\tau) (g^2/2) T},
\en
where 
\eq{ct}
c(\tau)=8\pi \int_\la^\infty e^{-\eps r^2} e^{-\tau r} dr.
\en
\el
\proof 
Let 
$
S_\eps^{\rm ren}
=S_\eps^{\rm OD}+Y+Z$, where 
\begin{align*}
Y&=
2 \int_{-T}^T \lk 
\int_{s}^{[s+\tau]} \nabla\varrho_\eps(B_t-B_s,t-s) dB_t \rk ds,\\
Z&=
-2 \int_{-T}^T \varrho_\eps(B_{[s+\tau]}-B_s,[s+\tau] -s)ds.
\end{align*}
It is established in \cite[Lemma 2.10]{ghl14} that 
\eq{k1}
\Ebb[e^{\alpha Y}]\leq e^{\alpha^2 T b_1}
\en
with some constant $b_1$. 
We estimate 
$\Ebb[e^{\alpha Z}]$. Straightforwardly there exists a constant 
$M>0$ such that 
$|\varrho_\eps(B_T-B_s,T-s)|\leq 
|\varrho_\eps(0,T-s)|<M$ for all $T$, and 
\begin{align*}
\varrho_\eps(0,T-s)\leq \half e^{-\la |T-s|}.
\end{align*}
Then we have 
\begin{align}
|Z|&\leq 
2\int_0^{2T} du \varrho_\eps(0,u)
=
2\lk\int_0^1+\int_1^{2T}\rk du \varrho_\eps(0,u)\non \\
&\label{k2}
\leq
2M+ \int_1^{2T} du \frac{1}{u} =
2M+\log(2T)-1.
\end{align}
Finally we can compute 
$S_\eps^{\rm OD}$. 
We have 
\begin{align}
|S_\eps^{\rm OD}|
&\leq 
2\int_{-T}^{T-\tau}
ds\int_{s+\tau}^T dt
\int_{|k|\geq\la}
\frac{1}{2\omega(k)}
e^{-\eps|k|^2}
e^{-\omega(k)|t-s|}
dk\non \\
&\label{k3}=
4\pi\int_\la^\infty 
e^{-\eps r^2}
\frac{e^{-\tau r}}{r}
\lk e^{-(2T-\tau)r}-1+(2T-\tau) r\rk 
dr\leq c(\tau) T.
\end{align}
Then  bound \kak{mmm2} follows from \kak{k1},\kak{k2}, \kak{k3} and 
the Schwarz inequality 
$
\Ebb[e^{(g^2/2)(S_\eps^{\rm OD}+Y+Z)}]\leq
\Ebb[e^{g^2 Y}]^\han 
\Ebb[e^{g^2(S_\eps^{\rm OD}+Z)}]^\han
$.
\qed
\begin{remark}
{\rm 
Constants $b$ and $c$ given in Lemma \ref{muri} also depend on $\tau$.
See \cite[Lemma 2.8, Lemma 2.10, (2.36)]{ghl14}.
}\end{remark}

Now we state the key lemma. 
\bl{main}
Let  $b>0$ and $c(\tau)$ be those in Lemma \ref{muri}. 
Then 
\eq{main1}
\left|
\frac{\El\ZZZ}{g^2}+\varrho_\eps(0,0)\right|
\leq \half(g^2 b+\half c(\tau)).
\en
\el
\proof
By Lemmas \ref{fkf} and \ref{fkf2} we have 
\eq{gse}
\El=-\lim_{T\to\infty}\frac{1}{2T}\log 
 \Ebb \left[
e^{\frac{g^2}{2}(S_\eps^{\rm ren}+4T\varrho_\eps(0,0)) }\right].
\en
We then have 
\begin{align*}
\El\ZZZ
=-g^2\varrho_\eps(0,0)-
\lim_{T\to\infty}\frac{1}{2T}\log 
\Ebb \left[
e^{\frac{g^2}{2}S_\eps^{\rm ren} }\right]
\end{align*}
Hence 
\begin{align*}
&|\El\ZZZ+g^2\varrho_\eps(0,0)|
\leq
\lim_{T\to\infty}\frac{1}{2T}\log 
 \Ebb \left[
e^{\frac{g^2}{2}S_\eps^{\rm ren} }\right].
\end{align*}
By Lemma \ref{muri} we can obtain  \kak{main1}. 	
\qed
We now state the main theorem in this paper. 
\bt{1}
It follows that 
\eq{main2}
\lim_{g\to 0}
\frac{\El\ZZZ}{g^2}=\ren 
\en and 
\eq{2}
\lim_{\eps\downarrow 0}\left|
{\El\ZZZ}-g^2 \ren 
\right|<\infty.
\en
\et
\proof
By Lemmas \ref{g} and \ref{main} we see that 
\eq{mainmain}
\left|
\frac{\El\ZZZ}{g^2}-\E\right|
\leq \half(g^2 b+\half c(\tau)).
\en
Take $g\to0$. We have
\eq{su}
\lim_{g\to 0}
\left|\frac{\El\ZZZ}{g^2}-\ren\right|\leq \frac{1}{4}c(\tau)
\en 
holds for arbitrary $\tau>0$. 
$\lim_{\tau\to\infty}c(\tau)=0$ implies 
\kak{main2}.
Furthermore 
\kak{2} can be derived from \kak{mainmain} and the fact 
 $\lim_{\eps\downarrow 0} c(\tau)<\infty$.
 \qed

\appendix
\section{Existence of the ground state}
For the self-consistency of the paper we show the uniqueness and the existence of ground state of $H_\eps(0)$. 
The proof mentioned below is taken from \cite[Lemma 2.9]{hir15}.  
Let 
$\grt=e^{-TH_\eps(0)}\one_\fff /\|e^{-TH_\eps(0)}\one_\fff \|$ 
%$ is a sequence converging to a ground state. 
and  $\gamma(T)=(\one_\fff , \grt)^2$, i.e., 
\eq{3-1}
\gamma(T)=\frac{(\one_\fff , e^{-TH_\eps(0)}\one_\fff )^2}{(\one_\fff , e^{-2TH_\eps(0)}\one_\fff )}.
\en

\bp{ex2}
For all $\eps>0$ and $\la>0$, 
$H_\eps(0)$ has the ground state and it is unique.
\ep
\proof
The uniqueness follows from the fact that 
$e^{-tH_\eps(0)}$ is positivity improving. 
It remains to show the existence of ground state. 
The useful criteria is as follows. 
There exists a ground state of $H_\eps(0)$ 
if and only if $\d \lim_{T\to\infty}\gamma(T)>0$ \cite{lms02}.
Thus it is enough to show 
$\d \lim_{T\to\infty}\gamma(T)>0$. 
By Lemma \ref{fkf} 
we have 
$$\gamma(T)=\frac{\lk
\Ebb[e^{\frac{g^2}{2}\int_0^T dt\int_0^T ds
W_\eps}]\rk^2}{
\Ebb[e^{\frac{g^2}{2}\int_{-T}^T dt\int_{-T}^T dsW_\eps}]}.$$
Here
$W_\eps=W_\eps(B_t-B_s, t-s)$. 
By the reflection symmetry and the Markov property of the Brownian motion 
we have 
$$\gamma(T)=\frac{\Ebb[e^{\frac{g^2}{2}
\int_{-T}^T dt\int_{-T}^T dsW_\eps-g^2\int_{-T}^0dt\int_0^T W_\eps}]}
{
\Ebb[e^{\frac{g^2}{2}\int_{-T}^T dt\int_{-T}^T dsW_\eps}]}.$$
By estimating $\int_{-T}^0dt\int_0^T W_\eps$ straightforwardly,  
we have 
\eq{big}
\gamma(T)\geq 
\exp\lk -g^2\int_\BR \kp \frac{ e^{-\eps|k|^2}}{\omega(k)^3}dk
\rk >0
\en
for all $T>0$. Note that $\la>0$. 
Then the proposition follows. 
\qed

\noindent
{\bf Acknowledgments:}
The author acknowledges support of Challenging Exploratory Research 15K13445 from JSPS.

{%\footnotesize

}

\end{document}